\documentclass{optica-article}

\journal{opticajournal} % for journals or Optica Open

\articletype{Research Article}

\usepackage{lineno}
\usepackage{graphicx}
\usepackage{cleveref} 
\Crefname{equation}{Eq.}{Eqs.}
\usepackage{capekCommands}
\usepackage{subcaption}

\usepackage[]{hyperref}
\hypersetup{
    colorlinks,
    linkcolor={red!50!black}, 
    citecolor={blue!50!black},
    urlcolor={blue!20!black}
}

\newacro{QCQP}[QCQP]{quadratically constrained quadratic program}
\newacro{VSWF}[VSWF]{vector spherical wave function}

\begin{document}

% \title{Limits of Optical Force and Stiffness \\ in the Dipole Regime}
\title{Physical Bounds on Optical Micromanipulation: Maximal Stiffness in the Dipole Regime}

\author{Martin Zlabek\authormark{1,*}, Jakub Liska\authormark{1,2}, Lukas Jelinek\authormark{1}, and Miloslav Capek\authormark{1}}

\address{\authormark{1}Czech Technical University in Prague, Technická 2, 166 27 Praha 6, Czech Republic}

\address{\authormark{2}Polytechnique Montréal, 2500 Chemin de Polytechnique, Montréal, QC H3T 1J4, Canada}

\email{\authormark{*}zlabemar@fel.cvut.cz} %% email address is required; see note below about the corresponding author designation

% use {abstract*} to suppress the copyright line. Copyright information will be added in production

\begin{abstract*}
Optical trapping and micromanipulation rely on carefully shaped electromagnetic fields to exert precise forces and torques on microscopic particles. Despite their widespread application in biology and nanotechnology, the absolute physical limits of trapping performance, specifically the maximum achievable optical force and trap stiffness, have not yet been rigorously quantified. This work establishes a general theoretical framework to determine these fundamental bounds in the dipole approximation. By relating the optical force and stiffness to a local Taylor expansion of the electromagnetic field at the particle location, we formulate the performance limit as a solution to a \acl{QCQP}. To evaluate these bounds, we employ two complementary approaches. First, we utilize a complete basis of \aclp{VSWF} to determine the absolute theoretical limits of optical force and stiffness permitted by Maxwell's equations in free space, revealing Pareto-optimal trade-offs between stable confinement and directional force. Second, we introduce an aperture-based formulation that restricts the incident fields to those realizable by finite planar apertures. This yields device-consistent bounds directly applicable to experimental setups which  rely mostly on electromagnetic beams. The finding that optimized aperture fields can outperform standard Gaussian beams by removing the severe axial bottleneck is particularly important. By comparing these two regimes, we identify the specific spatial modes that contribute to stable trapping and quantify the performance trade-offs inherent to physical beam shaping. This dual framework provides provably optimal bounds for power-normalized optical tweezers and serves as a rigorous benchmark for evaluating realistic beam designs.
\end{abstract*}

\section{Introduction}
Optical tweezers have become an essential tool for the manipulation of microscopic particles by using the radiation pressure and gradient forces exerted by focused optical beams \cite{1986_Ashkin_OL}. Their application spans diverse fields, from cell biology to quantum optics \cite{2021_Bustamante_NatRev,2021_Kaufman_NatPhys}, offering tools to apply small forces in a contact-free manner \cite{2004_Neuman_RSI, 2021_Gieseler_etalOpticalTweezers}. The physical foundation of optical trapping is well understood through Maxwell's equations and the dipole approximation \cite{Zangwill_Modern_Electrodynamics}. However, the fundamental limits of achievable trapping forces and stiffness remain largely unexplored. The developed framework offers a means to evaluate how far current state-of-the-art experimental setups are from reaching their fundamental limits. 

This work builds on the broader program of fundamental bounds in electromagnetics, which originated in antenna-performance limits~\cite{GustafssonTayliEhrenborgEtAl_AntennaCurrentOptimizationUsingMatlabAndCVX},  was later extended to quadratic metrics as well as convex-optimization formulations~\cite{2022_Chao_PhysicalLimitsEM,Liska_etal_FundamentalBoundsEvaluation} and subsequently applied to the study of electromagnetic forces such as magnetic levitation~\cite{Liska_DiplomaThesis}, magnetic confinement~\cite{Liska_PerformanceBoundsOfMagneticTrapsForNeutralParticlesPRA}, and dielectrophoresis~\cite{Zlabek_2025} using local approximations. In the optics context, related progress has already been made in formulations based on generalized eigenvalue problems, increasingly general linear and quadratic constraints, and aperture-constrained focusing bounds~\cite{Schabetal_2022_UpperBoundsOnFocusing}. We use the same QCQP framework and the accompanying fundamental-bounds implementation developed in~\cite{Liska_etal_FundamentalBoundsEvaluation}.

Recent studies have approached optimal trapping by considering specific beam types, such as Gaussian or vortex beams \cite{2009_Albaladejo_PRL, 2017_Gao_LSA}. Other works employ rigorous computational techniques such as the generalized Lorenz-Mie theory \cite{2014_Nieminen_JQSRT_OpticalTweezersTheoryAndModelling}. While these approaches are highly valuable for analyzing existing optical setups, they evaluate specific configurations and do not quantify the fundamental limits imposed purely by the laws of electromagnetism.

This work addresses this gap by developing a general framework for determining the absolute physical bounds on optical tweezing and trapping performance in a homogeneous lossless background medium. Using a local Taylor expansion of the electromagnetic field at the particle location~\cite{Liska_2025_IntSocOP}, we cast the optical force and trap stiffness as quadratic forms of the field expansion coefficients. We then formulate the search for maximum performance as an optimization problem under power or energy normalization and force balance constraints. By leveraging convex optimization through \acp{QCQP} \cite{BoydVandenberghe_ConvexOptimization}, this method guarantees globally optimal results.

To comprehensively evaluate these limits, this optimization framework is applied to two distinct regimes. First, we expand the incident field using regular \acp{VSWF}. As this basis spans the entire source-free radiation space, it gives the absolute fundamental bound on force and stiffness permitted by Maxwell's equations. Second,  an aperture-based formulation is introduced. Experimental optical tweezers rely on finite-aperture lenses \cite{2004_Neuman_RSI} meaning the \ac{VSWF} limit might overestimate practically achievable performance. The aperture method restricts the optimization to only those fields realizable by a specific planar aperture. This provides a much tighter bound that closely reflects the behavior of realistic optical systems. Crucially, this formulation also establishes a common mathematical framework for comparing the fundamental limits with standard paraxial Gaussian beams which are frequently used to model laboratory experiments \cite{1996_Harada_OptCommun, 2012_NovotnyHecht_NanoOptics}.

By comparing the absolute \ac{VSWF} limits with the device-consistent aperture limits and conventional Gaussian profiles, we expose how specific spatial modes contribute to trapping efficiency and stability. This dual framework serves as a rigorous reference for evaluating realistic beam designs and optical micromanipulation strategies.

\section{Theory \label{sec:theory}}
Optical tweezers and optical traps are fundamentally the same technology, and the terms are often used interchangeably. Both use tightly focused laser beams to hold and manipulate microscopic dielectric particles \cite{2011_Volpe_CONTEMPT}. The term trap usually refers to a stable, motionless potential well in which the net optical force on the particle vanishes. Conversely, tweezers refer to the system's ability to move a dielectric object or counterbalance external forces, such as uniform fluid drag or gravity, effectively serving as an optical manipulator. It is commonly assumed that the external forces act uniformly across the trapping volume, lack spatial curvature, and do not contribute to the local force gradients. Consequently, the mathematical description of stability remains unified for both scenarios. For the sake of brevity, the remainder of this text will primarily utilize the word \textit{trap}, though the theoretical limits and principles derived herein apply equally to a tweezing configuration.

To investigate the performance limits of optical trapping systems, we model the interaction between an optical field and a sub-wavelength dielectric particle using the dipole approximation. In the Rayleigh regime, the optical force and its spatial derivatives are determined entirely by the local electric field and its gradients evaluated at the particle location.

\subsection{Optical Force and Stiffness Matrix}
We consider a time-harmonic electric field $\V{E}(\V{r})$ at a single angular frequency $\omega$ with time convention~$\T{exp} \left( - \T{i} \omega t \right)$. For an isotropic polarizable particle, the cycle-mean optical force evaluated in the dipole approximation is given by \cite{Zangwill_Modern_Electrodynamics}
\begin{equation}
\label{eq:force}
    \langle \V{F} \rangle = \dfrac{1}{2} \varepsilon \RE \left\{ \alpha^* (\V{E}^* \cdot \nabla) \V{E} + \alpha^* \V{E} \times (\nabla \times \V{E}) \right\},
\end{equation}
where $\varepsilon$ is the real-valued permittivity of the background and $\alpha$ is the complex-valued electric polarizability of the particle with units of volume. The force and electric field are evaluated at the position of the particle which can conveniently be placed at the origin of a coordinate system. To satisfy energy conservation, dynamic polarizability must include a radiation reaction correction~\cite{Tretyakov2003}
\begin{equation}
\label{eq:dynamic_pol}
    \alpha = \dfrac{\alpha_0}{1 - \T{i} \dfrac{k^3}{6\pi} \alpha_0},
\end{equation}
where $k$ is the wavenumber and~$\alpha_0$ is the static electric polarizability with the dimensions of volume. The complex nature of $\alpha$ captures the full linear electromagnetic response of the scatterer. Its real part determines the strength of the conservative gradient force that draws the particle toward regions of higher intensity. Its imaginary part gives rise to a non-conservative\footnote{A non-conservative force does not allow us to define the potential.} scattering force that pushes the particle in the direction of energy flow~\cite{Zangwill_Modern_Electrodynamics}. A convenient dimensionless parameterization of the static polarizability is obtained by separating its magnitude and phase as
\begin{equation}
\alpha_0 = \zeta \dfrac{6\pi}{k^3} \T{e}^{\T{i}\phi}.
\end{equation}
Here, dimensionless parameter $\zeta$ serves as a normalized electrical volume, and phase angle $\phi$ isolates the intrinsic material dissipation before radiation losses are applied via \Cref{eq:dynamic_pol}. %For a passive material, the physically valid phase is restricted to $\phi \in [0,\pi]$, with $\phi=0$ representing the standard lossless reference case and $\phi=\pi/2$ representing a purely dissipative particle. The endpoint $\phi=\pi$ corresponds to a purely real polarizability of the opposite sign.
For a passive material, the physically valid phase is restricted to $\phi \in [0,\pi]$, with $\phi \in \{ 0, \pi \}$ representing the lossless cases and $\phi=\pi/2$ representing a purely dissipative particle.

The mathematical expression~\Cref{eq:force} shows that the net force at the origin depends exclusively on the Cartesian components $p \in \{x, y, z\}$ of the local electric field $E_j (\V{0})$ and its first spatial derivatives $\partial_i E_j (\V{0})$ coupled with the material properties. Furthermore, the cross products can be algebraically simplified to reveal a much more compact expression. By evaluating the time-averaged optical force along an arbitrary Cartesian axis, the relation reduces to \cite{2004_Anatoly_Rolay}
\begin{equation}
\label{eq:force_component}
    \langle F_p \rangle = \dfrac{1}{2} \varepsilon \RE \left[ \alpha^* \sum \limits_j E_j^* \partial_p E_j \right].
\end{equation}
This component-wise formulation greatly simplifies subsequent algebra and clearly isolates the directional gradients required to evaluate the stability of the trap.

The fundamental objective of an optical trap is to stably confine a particle to a specified spatial coordinate. At this target location, the net time-averaged optical force must either vanish or counter-balance the external forces. At the same time, the surrounding optical field must exert a restoring force that actively opposes the displacement of the particle away from the trapping center. By applying a linear approximation to the local force field in the neighborhood of the equilibrium point, we can mathematically capture this restoring behavior. The spatial linearization of the optical force field leads to the stiffness matrix which governs local stability. It is nonetheless crucial to note that this formulation does not account for larger spatial displacements, nor can it accurately track restoring forces that exhibit strong non-linear dependencies farther from the trapping center.

\begin{figure}[t!]
    \centering
    % First Subfigure
    \begin{subfigure}{0.48\textwidth}
        \centering
        \includegraphics[width=\linewidth]{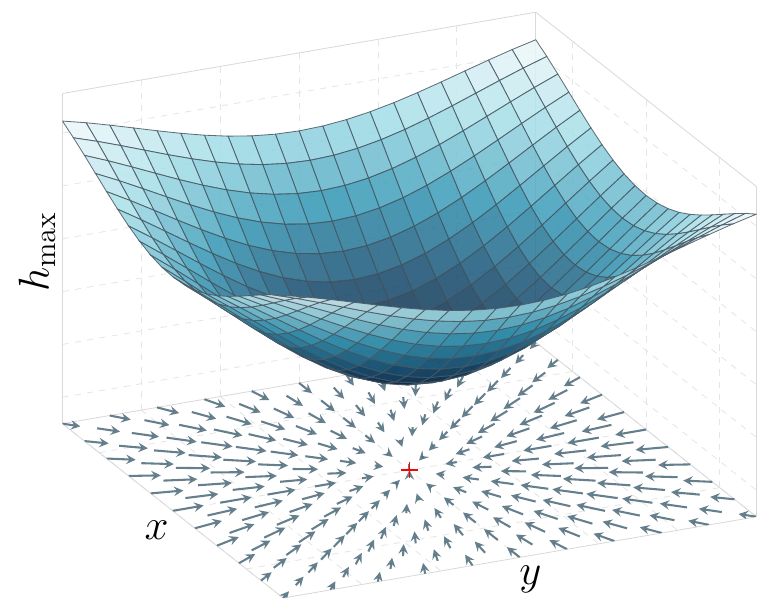}
        \caption{Isotropic trap}
        \label{fig:ideal_trap}
    \end{subfigure}\hfill 
    \begin{subfigure}{0.48\textwidth}
        \centering
        \includegraphics[width=\linewidth]{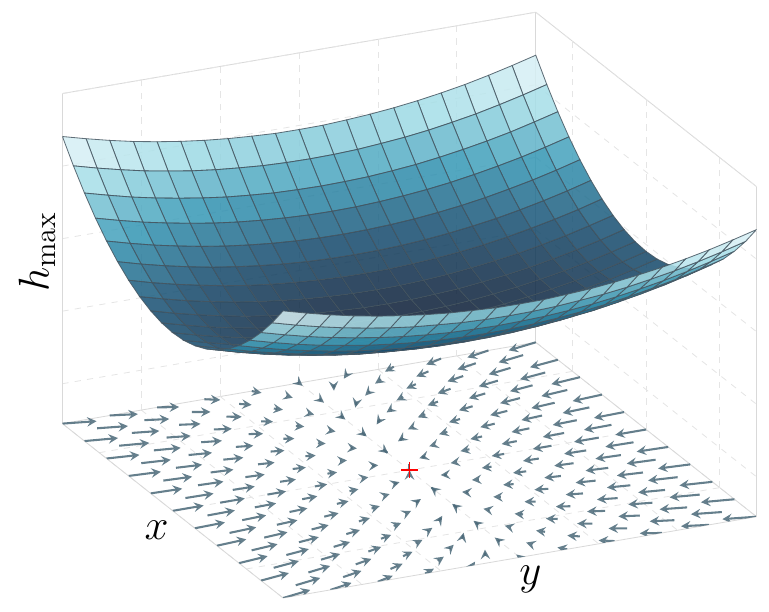}
        \caption{Anisotropic trap}
        \label{fig:non_ideal_trap}
    \end{subfigure}
        \caption{Comparison of the maximum eigenvalue of the stiffness matrix for (a) a trap with optimal symmetric stiffness and (b) a non-symmetric trap lacking isotropic stiffness. The trap is characterized by its force field, which must vanish at the trapping center, and the stiffness matrix, which must be negative definite to enforce trap stability.}
    \label{fig:trap_comparison}
\end{figure}

The stiffness matrix components are defined by the spatial derivatives of the force
\begin{equation}
\label{eq:stiffness}
    \langle H_{pq} \rangle = \partial_p \langle F_q \rangle.
\end{equation}
Evaluating this stiffness requires the first $\partial_i E_j$ and second $\partial_i \partial_j E_k$ spatial derivatives of the electric field at the trapping center. The local approximation approach is similar to the computation of the performance limits on quasi-static forces~\cite{Liska_DiplomaThesis,Liska_PerformanceBoundsOfMagneticTrapsForNeutralParticlesPRA,Zlabek_2025}.
Eigenvalues of the stiffness matrix quantify the strength of the restoring force along the principal directions. A stable trap requires this matrix to be negative definite, so all of its eigenvalues must be negative. To characterize the overall trap strength with a single scalar measure, we use the largest eigenvalue
\begin{equation}
    h_{\max} = \max\{\operatorname{eig}(\langle \V{H} \rangle)\},
\end{equation}
which corresponds to the weakest restoring direction and thereby determines the limiting confinement strength of the trap. For isotropic traps, all eigenvalues are equal, so $h_{\max}$ provides the same measure of confinement as any diagonal component of $\V{H}$. This concept is depicted in \Cref{fig:trap_comparison} which contrasts the trap with the optimal symmetric stiffness featuring isotropic restoring forces and a trap that lacks such symmetry.

\section{Field Representation and Optimization}\label{sec:LocExpansion}

To find an optimal trap or tweezers, we need access to all possible field configurations in the vicinity of the trapping center. To that end, the incident electric field is represented as a linear combination of a chosen set of source-free Maxwell-consistent basis functions $\basisFcn_n(\V{r})$
\begin{equation}
\label{eq:fieldExpansion}
    \V{E}(\V{r}) = k \sqrt{Z} \sum_{n} a_n \, \basisFcn_n(\V{r}),
\end{equation}
where $Z$ is the wave impedance of the background medium, and $a_n$ are the complex expansion coefficients forming column vector $\M{a}$. Since Maxwell's equations and the spatial derivative operators are linear, the local field and its derivatives evaluated at the origin are linear forms of vector~$\M{a}$, specifically~\cite{Zlabek_2025}
\begin{equation}
    E_p = \M{E}_p \M{a}, \quad
    \partial_i E_p = \partial_i \M{E}_p \M{a} , \quad
    \partial_i \partial_j E_p = \partial_i \partial_j \M{E}_p \M{a}.
    \label{eq:fieldMatrices}
\end{equation}
% By restricting our analysis to the harmonic regime, we approximate the trapping potential as a three-dimensional quadratic well. In this regime, the trap stability is completely determined by the local field and its spatial derivatives up to the second order evaluated strictly at the trap center ($\V{r} = \V{0}$). Because these local field properties depend linearly on $\M{a}$, 
Substituting the expansion into the optical force \Cref{eq:force_component} and stiffness \Cref{eq:stiffness} naturally yields Hermitian quadratic forms. For arbitrary spatial direction $p$, the time-averaged force component is evaluated as
\begin{equation}
    \langle F_p \rangle = \M{a}^\herm \M{F}_p \M{a},
\end{equation}
where $\M{F}_p$ is a Hermitian matrix constructed from the basis functions and the complex polarizability $\alpha$. By the same exact mathematical principle, the stiffness matrix elements become
\begin{equation}
    \langle H_{pq} \rangle = \M{a}^\herm \M{H}_{pq} \M{a},
\end{equation}
where $\M{H}_{pq}$ denotes the matrix representations of the individual components of the stiffness operator.

Finding the optimal trap or tweezers is equivalent to identifying vector~$\M{a}$ that enforces the vanishing or counter-balancing force at the trapping center, and gives minimal maximum eigenvalue~$h_{\max}$  of the stiffness matrix. Minimizing eigenvalue~$h_{\max}$ directly is generally a non-smooth optimization problem. To cast this into a more tractable form, we restrict our search space to isotropic traps. By rotating the coordinate axes to align with the principal axes of the trap and actively enforcing a diagonal stiffness matrix with equal diagonal coefficients, all three eigenvalues become identical.
Under this isotropic assumption, the non-smooth minimax problem can be replaced by the minimization of the trace of the stiffness matrix. As both the optical force and the stiffness matrix components are Hermitian quadratic forms of $\M{a}$, the trace serves as a smooth quadratic objective. Furthermore, enforcing the isotropic conditions—setting off-diagonal stiffness components to zero and forcing diagonal components to be equal—generates a set of quadratic equality constraints. Consequently, the entire optimization process can be formulated as a \ac{QCQP}.

The primal optimization problem is inherently non-convex. To systematically find the fundamental limits, we instead formulate and solve its Lagrangian dual problem which is strictly convex by definition \cite{BoydVandenberghe_ConvexOptimization}. For the specific algebraic structure of the operators evaluated in this work, we observe strong duality meaning the duality gap is zero. This guarantees that the solution to the convex dual problem yields globally optimal bounds for the original physical system. In the context of this paper, the QCQP is solved by means of convex optimization~\cite{NocedalWright_NumericalOptimization} using a freely downloadable solver~\cite{FunBoPackage} described in~\cite{Liska_etal_FundamentalBoundsEvaluation}.

The last essential ingredient in forming a solvable optimization problem for the optimal trap is to limit the field amplitudes. This constraint can be based on any positive definite matrix $\M{P}$ which limits the field values via a relation
\begin{equation}
    \M{a}^\herm \M{P} \M{a} = P_0,
    \label{eq:posDefMat}
\end{equation}
where~$P_0$ is typically a cycle-mean power or energy.

In the particular case of isotropic spherical traps treated in \Cref{sec:VSWF}, the cycle-mean power incident on the trapped particle is used. For the single-aperture configurations treated in \Cref{sec:aperture}, the average value of the cycle-mean energy density per unit length in a plane is a more appropriate metric. Lastly, for finite volumes enclosed by radiating apertures treated in \Cref{sec:vsw_vs_ap}, the average value of the cycle-mean energy density per unit length over a sphere enclosing the trapped particle is used. Other possibilities exist, and their effect results in different amplitude normalization. The shape of the force field in the vicinity of the trapping point is given by the balance between optical force \Cref{eq:force_component} and stiffness \Cref{eq:stiffness}.

With the basis established, the specific problem of determining the fundamental limit on the performance of an optical trap (or tweezers) can be formulated. By taking advantage of the \ac{QCQP} structure derived above, the optimization minimizes the trace of the stiffness matrix subject to power normalization, force balance, and isotropy constraints
\begin{equation}
\label{eq:QCQP_vswf_stiffness}
    \begin{aligned}
        \min_{\M{a}} \quad & \T{Tr} \left( \left[ \M{a}^\herm \M{H}_{pq} \M{a} \right] \right) \\
        \text{s.t.} \quad & \M{a}^\herm \M{P} \M{a} = P_0 \\
        & \M{a}^\herm \M{F}_i \M{a} = F_{0,i}, \quad \text{for } i \in \{x, y, z\} \\
        & \M{a}^\herm (\M{H}_{ii} - \M{H}_{jj}) \M{a} = 0, \quad \text{for all } i \neq j \\
        & \M{a}^\herm \M{H}_{ij} \M{a} = 0, \quad \text{for } i \neq j.
    \end{aligned}
\end{equation}
where $\T{Tr}$ denotes a trace and $F_{0,i}$ represents the target optical force which is either zero for static confinement or equal and opposite to any macroscopic external forces being counter-balanced. For an isotropic Hessian, the diagonal entries coincide and, thus, determine the eigenvalues of the stiffness matrix directly. 

Directly solving~\cite{Liska_etal_FundamentalBoundsEvaluation} the problem in \Cref{eq:QCQP_vswf_stiffness} is computationally demanding and prone to numerical instabilities. To ensure the numerical convergence of the \ac{QCQP} solver~\cite{FunBoPackage}, it is advantageous to eliminate redundant degrees of freedom. To achieve this, we first recognize that for any monochromatic wave, the electric field vector at a specific spatial point traces an ellipse lying entirely within a two-dimensional plane. By projecting the expansion coefficients into the nullspace of the operator governing the out-of-plane field component, we restrict the local electric field to this plane. To guarantee that the true global optimum is preserved, this polarization plane must be systematically aligned with the principal axes of the trap symmetry.

Second, we isolate the degrees of freedom that actively contribute to the trapping force field. We define the local feature matrix $\M{A}$ to extract the field and its relevant derivatives at the trapping center
\begin{equation}
    \M{A} = \left[ \M{E}_p^\trans, \left(\partial_i \M{E}_p\right)^\trans, \left(\partial_i \partial_j \M{E}_p\right)^\trans \right]^\trans .
\end{equation}
Following the definition \Cref{eq:fieldMatrices}, matrix $\M{A}$ has 30 lines and as many columns as the number of basis functions $\V{\psi}$. Vectors~$\M{a}$ in the nullspace of this matrix create no field, no force, and no stiffness, and are of no use for a trap or a tweezers in linear force approximation\footnote{In general, higher-order approximations would require higher-order derivatives.}. The rows of matrix~$\M{A}$ make explicit which field derivatives are relevant for the optimization. We then construct a generalized eigenvalue problem \cite{Liska_2025_IntSocOP}
\begin{equation}
\label{eq:gevp_reduction}
    \M{A}^\T{H} \M{A} \V{v}_m = \lambda_m \M{P} \V{v}_m,
\end{equation}
where it is important to remember that the rank of matrix~$\M{A}^\T{H} \M{A}$ is at most $30$. Modes associated with zero eigenvalues~$\lambda_m$ represent dark modes consuming power (or energy) without contributing to the trapping metrics. All these modes can be removed from the degrees of freedom of the optimization problem described by~\Cref{eq:QCQP_vswf_stiffness}.

To simultaneously eliminate non-contributing dark modes and confine the local electric field to the chosen two-dimensional polarization plane, we construct compact transformation matrix $\M{T}$ spanning the relevant degrees of freedom collected in vector~$\widetilde{\M{a}}$. This matrix maps a reduced set of active coefficients back to the original full space via 
\begin{equation}
    \M{a} = \M{T} \widetilde{\M{a}}   
\end{equation}
Applying this congruence transformation to all relevant matrices (\textit{e.g}., $\M{T}^\herm \M{H}_{pq} \M{T}$) allows us to solve the exact \ac{QCQP} formulated in \Cref{eq:QCQP_vswf_stiffness} entirely within this compressed subspace. Solving this dimensionally reduced problem yields the absolute theoretical upper bound for trapping stiffness permitted by Maxwell's equations and the degrees of freedom used.

\section{Vector Spherical Waves \label{sec:VSWF}}

To determine the absolute physical limits of optical trapping in free space, we must optimize the complete set of valid electromagnetic fields. The regular \acp{VSWF} provide a natural and mathematically complete basis for any source-free field in a homogeneous medium \cite{2016_Kristensson_Scattering}.

To establish these fundamental limits, the generic basis functions $\basisFcn_n(\V{r})$ in the local field expansion from \Cref{eq:fieldExpansion} are explicitly chosen as the regular \acp{VSWF}, denoted as $\V{u}^{(1)}_\nu(k \V{r})$. In this context, generic index $n$ is replaced by multi-index $\nu$ which captures the spherical harmonic degree $l$, azimuthal order $m$, and polarization state. A detailed mathematical definition of these wave functions and associated notation is provided in \Cref{apx:VSWF}.

Because the harmonic trap is fully characterized by spatial derivatives up to the second order, the relevant electromagnetic degrees of freedom at the origin are fundamentally limited to multipole moments up to the degree $l = 3$ (octupole). The \ac{VSWF} expansion perfectly aligns with this requirement. regular spherical waves of order $l$ scale as $r^l$ near the origin, so all modes with $l>3$ and their derivatives up to second order vanish at the origin \cite{Liska_2025_IntSocOP}. To bound the field amplitudes and evaluate the fundamental limits under a constrained power budget, we normalize the \ac{QCQP} from \Cref{eq:QCQP_vswf_stiffness} by the cycle-mean inward power converging on the trap ($P_0 = P_\T{in}$). This specific metric is chosen because it robustly limits the total electromagnetic energy flux into the focal volume from all directions. As detailed in \Cref{apx:VSWF}, the spherical orthogonality of the \ac{VSWF} basis dramatically simplifies this power evaluation. The power normalization matrix reduces to scaled identity matrix~$\M{I}$, specifically, $\M{P} = \M{I} / 8$.

\subsection{Fundamental Limits of Optical Trapping \label{sec:results}}

\begin{figure}[t!]
    \centering
    \includegraphics[width=\linewidth]{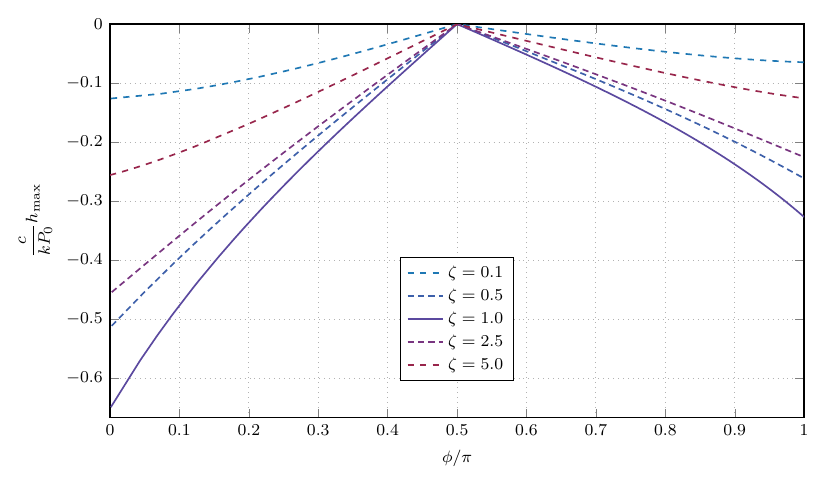}
    \caption{Absolute upper bounds of dimensionless trap stiffness $(c \, h_{\max})/ (k P_0)$ evaluated as a function of polarizability phase $\phi$ for various normalized particle volumes $\zeta$.}
    \label{fig:trace_plot}
\end{figure}

By evaluating the optimal \ac{VSWF} fields over a wide range of the normalized polarizability parameters $\zeta$ and $\phi$ defined in \Cref{eq:dynamic_pol}, we map the global bounds of isotropic trapping. The resulting physical limits are visualized in \Cref{fig:trace_plot}. The ideal confinement is quantified by dimensionless quantity $c\, h_{\max}/(kP_0)$, where $c$ is the speed of light, which captures the weakest restoring direction and, thus, sets the limiting trap strength.

For small values of $\zeta$, the particle operates in a strict Rayleigh regime where this optimal stiffness scales linearly with the particle volume. However, as intrinsic material dissipation increases ($\phi \to \pi/2$), performance degrades. All stiffness bounds strictly converge to zero at $\phi = \pi/2$. While it is well established that a single focused beam in a lossless mediums cannot trap a purely dissipative particle within the linear force approximation due to the lack of a gradient force~\cite{Ashkin_1983_Optical_Earnshaw}, this result confirms that no complex multi-beam configuration can overcome this limitation. For $\phi = \pi/2$, the real part of the dynamic polarizability vanishes, leaving only the non-conservative scattering force \cite{Zangwill_Modern_Electrodynamics}. In a source-free lossless background medium, the divergence of this non-conservative force field is strictly zero, meaning the eigenvalues of its corresponding stiffness matrix must be zero. While such a field might provide a restoring force along certain spatial axes, the zero trace dictates that it must be proportionally anti-restoring along the others. Consequently, it is mathematically impossible to achieve the fully bounded three-dimensional restoring curvature (a negative definite stiffness matrix) required for a stable trap without the conservative gradient response.

As normalized volume $\zeta$ approaches unity, the constraints imposed by the radiation reaction become dominant, and the achievable trap stiffness saturates. The maximum theoretical trap stiffness occurs exactly at $\zeta = 1$. This limit emerges from the interplay between the conservative gradient force, which is proportional to $\RE(\alpha)$, and the non-conservative scattering force, which is proportional to $\IM(\alpha)$. To satisfy the equilibrium constraint of zero net force at the trap center ($\langle \V{F} \rangle = \V{0}$), the optical field must be structured so that the gradient force balances the radiation pressure. Analytically, the real part of dynamic polarizability $\RE(\alpha)$ reaches its global maximum at $\zeta = 1$, providing the strongest possible gradient response. Substituting the static polarizability into the dynamic polarizability equation (\Cref{eq:dynamic_pol}) yields $\RE(\alpha) \propto \zeta \cos\phi / (1 + 2\zeta\sin\phi + \zeta^2)$. This functional form reveals a strict inversion symmetry, $\RE(\alpha_{\zeta}) = \RE(\alpha_{1/\zeta})$, dictating that the conservative gradient response strictly mirror itself for volumes scaled inversely across the $\zeta = 1$ threshold. For volumes exceeding $\zeta = 1$, this gradient response symmetrically decreases while $\IM(\alpha)$ (radiation damping) grows monotonically. The incident field must therefore balance an increasing scattering force with a decreasing gradient force.

\begin{figure}[t!]
    \centering
    \includegraphics[width=\linewidth]{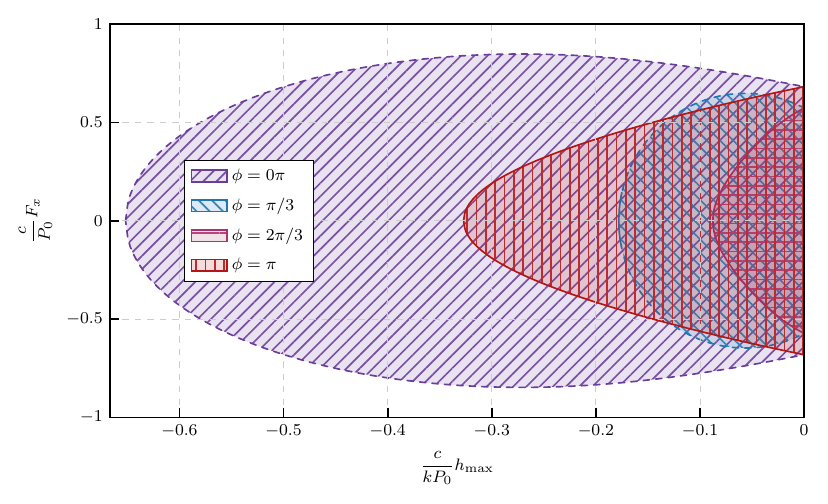}
    \caption{Fundamental trade-off between dimensionless trap stiffness $(c h_{\max})/( k P_0)$ and directional optical force $(c F_x)/(P_0)$. For fixed normalized particle volume $\zeta = 1.0$ and various polarizability phases $\phi$, the shaded areas denote all physically accessible trapping states. The solid boundaries mark the absolute Pareto optimal limits.}
    \label{fig:pareto_plot}
\end{figure}

While \Cref{fig:trace_plot} establishes the bounds for a stationary trap under a strict zero-net-force constraint, practical optical micromanipulation frequently requires exerting a controlled translational force on the particle. To quantify the limits of this dynamic regime, we evaluate the complete feasible region of trapping states, defined by the trade-off between the trap stiffness and the maximum achievable directional force. \Cref{fig:pareto_plot} illustrates these regions by plotting dimensionless optical force component $c F_x / P_0$ against dimensionless trap stiffness $c\, h_{\max} / (k P_0)$ for a fixed normalized volume $\zeta = 1.0$.

The shaded areas in \Cref{fig:pareto_plot} represent all physically accessible field configurations, while the solid boundary lines mark the Pareto-optimal frontiers demonstrating the inherent trade-off between trapping stability and directional force. For a lossless particle ($\phi = 0$), maximizing the stiffness (the leftmost extremum) yields a suboptimal pulling force. If the stability requirements of the trap are relaxed, the optical field can be restructured to exert a stronger directional force, culminating in a distinct performance peak. However, as intrinsic particle losses increase ($\phi > 0$), the feasible region shrinks towards the origin. This graphically confirms that energy dissipation degrades the capacity of the optical field to simultaneously maintain high stiffness and deliver strong translational forces.

To systematically construct these Pareto frontiers, multiobjective optimization is resolved as a sequence of constrained \acp{QCQP}. We first establish the theoretical baseline by solving for the minimum trace of the stiffness matrix while relaxing other constraints. Subsequently, we relax the zero-force constraint along a single translational axis and introduce the stiffness trace as a bounding equality constraint. By sweeping this target trace from its optimal minimum towards zero and iteratively maximizing directional force component $F_x$ at each step, we obtain the complete outer boundary of physically accessible field configurations.

\section{Axisymmetric Apertures\label{sec:aperture}}

While the \ac{VSWF} basis establishes the absolute physical bounds permitted by Maxwell's equations in free space, practical optical tweezers are constrained by the finite numerical aperture of their focusing lenses. To evaluate device-consistent limits, we restrict the available electromagnetic degrees of freedom to those generated by finite axisymmetric apertures.

Time-harmonic electric field $\V{E} (\rho, z > 0)$ of a single circular aperture of radius $R$ lying at $z = 0$ can be formulated by combining the spatial Hankel transform with the angular spectrum decomposition~\cite{2012_NovotnyHecht_NanoOptics}, see~\Cref{appAperture} for details. The field is rotationally symmetric about the axis of the aperture and linearly polarized. For arbitrary polarization, two orthogonally polarized apertures have to be considered at the same location. The electric field is decomposed into its transverse~$\V{E}_\bot$ part and longitudinal part~$E_z$. Assuming the transverse part of the aperture field to be linearly polarized along unit vector $\hat{\M{e}}$, this yields
\begin{equation}
    \V{E}_\bot (\rho, z) = \hat{\M{e}} \int \limits_{0}^{\infty} \int \limits_{0}^{R} E_\bot(\rho') \T{J}_0(k_\rho \rho')  \T{J}_0(k_\rho \rho) \T{e}^{\T{i} k_z z} \rho'  \, \T{d} \rho' k_\rho \T{d} k_\rho,
    \label{eq:propagated_field}
\end{equation}
where $\rho$ and $\rho'$ are the radial coordinates in the observation and aperture planes, respectively, $k_\rho$ is the transverse wavenumber, $k_z = \sqrt{k^2 - k_\rho^2}$ (with $\T{Im} \left\{ k_z \right\} > 0$) is the longitudinal wavenumber, and $\T{J}_n$ is the $n$-th order Bessel function. The longitudinal component of the field is uniquely determined by the requirement of vanishing divergence at every point. 

To integrate this propagation model into the optimization framework, we expand aperture field~$E_\bot(\rho)$ using a set of basis functions $\Phi_n(\rho)$ strictly confined to interval $\rho \in [0, R]$. This truncation reflects the physical reality of experimental optical systems. We assume that aperture radius $R$ is large enough to capture the entire incident beam profile which means that the field naturally decays to zero at the boundary. Any marginal field that extends beyond $R$ is physically blocked by the opaque housing of the objective lens. Therefore, it is both mathematically convenient and physically rigorous to enforce a hard boundary condition, $E_\bot (R) = 0$, at the edge of the aperture. Substituting this representation into the propagation integral~\Cref{eq:propagated_field} and using~\Cref{appAperture} identifies the global basis functions $\basisFcn_n(\rho, z)$
\begin{equation}
\begin{aligned}
    \basisFcn_{\bot,n}(\rho, z) &= \hat{\M{e}} \int \limits_{0}^{\infty} I_n (k_\rho) \T{J}_0(k_\rho \rho) \T{e}^{\T{i} k_z z} k_\rho \, \T{d} k_\rho, \\
    \basisFcn_{z,n}(\rho, z) &= - \T{i} \hat{\V{r}}_\bot  \cdot \hat{\M{e}} \int\limits_0^\infty \dfrac{k_\rho}{k_z} I_n (k_\rho) \T{J}_1(k_\rho \rho) \T{e}^{\T{i} k_z z} k_\rho \, \T{d} k_\rho, \\
    I_n (k_\rho) &= \int \limits_{0}^{R} \Phi_n(\rho) \T{J}_0(k_\rho \rho) \rho \, \T{d} \rho,
    \label{eq:propagated_basis}
\end{aligned}
\end{equation}
into which the electric field is expanded via~\Cref{eq:fieldExpansion}. The transverse position vector in the aperture plane is denoted by~$\V{r}_\perp=x\hat{\V{x}}+y\hat{\V{y}}$ and~$ \hat{\V{r}}_\bot = \V{r}_\bot / \left| \V{r}_\bot \right|$.

Depending on the required flexibility, different choices can be considered for the aperture basis functions $\Phi_n(\rho)$. For applications such as topology optimization, where a sparse overlap matrix and highly localized control over arbitrary field profiles are desirable, a local basis composed of piecewise-linear rooftop (tent) functions might be preferable. However, for the rigorous global optimization of typical trap profiles, a global Fourier-Bessel series consisting of orthogonal Bessel modes offers significant computational advantages. Therefore, we expand the aperture field as
\begin{equation}
\begin{aligned}
    \Phi_n(\rho) &= \T{J}_0(\alpha_n \rho/R), \quad n = 1, \dots, N, \\
    I_n (k_\rho) &=  R^2  \dfrac{ \alpha_n \T{J}_0(k_\rho R) \T{J}_1 (\alpha_n)}{\alpha_n^2 - (k_\rho R)^2},
\label{eq:BesselBasis}
\end{aligned}
\end{equation}
where $\alpha_n$ represents the $n$-th root of Bessel's function~$\T{J}_0$. Although \Cref{eq:BesselBasis} appears to have a pole at $k_\rho R = \alpha_n$, this singularity is removable
\begin{equation}
    \lim_{k_\rho R \to \alpha_n} I_n(k_\rho) = \frac{R^2}{2}\,\T{J}_1^2(\alpha_n).
\end{equation}
This basis naturally satisfies the hard boundary condition at the edge of the aperture. To validate the aperture framework, its results are compared with the analytical paraxial model of the Gaussian beam. This comparison showed that the present formulation accurately captures the beam expansion and associated spatial phase variation in the regime where the paraxial model is expected to hold.

A critical aspect of formulating the \ac{QCQP} for finite apertures is selecting a robust amplitude bounding metric. If the optimization is normalized strictly by the active propagating power radiated into the far-field, the algorithm tends to heavily exploit the evanescent spectrum. While evanescent modes are valid solutions to Maxwell's equations, exploiting them enables the optimizer to synthesize highly localized sources with no cost in active power. Because experimental optical traps are typically formed by propagating modes rather than extreme near-field reactive components, we seek to avoid this behavior.

To suppress the generation of these evanescent artifacts, we bound field amplitudes by the time-averaged electromagnetic energy density integrated across the aperture plane at $z=0$
\begin{equation}
    \label{eq:aperture_intensity}
    W_\T{ap} = \frac{1}{4} \int \limits_\T{ap}  \left( \varepsilon |\V{E}|^2 + \mu |\V{H}|^2 \right) \, \T{d} S.
\end{equation}
Employing spectral integration, see~\Cref{appAperture} and~\Cref{eq:WcycleMean}, the matrix form of the positive definite constraint~\Cref{eq:posDefMat} is obtained.

It should be noted that normalizing the system by the active propagating power radiated into the far-field represents a valid alternative approach. However, to prevent the optimizer from exploiting evanescent waves (which carry zero cycle-mean active power) to artificially inflate trap stiffness, the active power formulation requires the strict limitation of basis mode count $N$ to purely propagate spatial frequencies ($\alpha_N / R \le k$). Since the number of available propagating modes is fundamentally dictated by the electrical size of the aperture ($kR$), this power-bounded approach is only viable for large apertures that provide sufficient degrees of freedom to effectively shape the beam. For a generalized computational framework that must accommodate varying numerical apertures, the energy-based metric from \Cref{eq:aperture_intensity} is significantly more robust.

With the continuous field successfully discretized, the optical force, trap stiffness, and aperture intensity are recast into Hermitian forms of the expansion coefficients collected in state-space vector $\M{a}$. The system translates into the characteristic Hermitian matrices $\M{F}_p$, $\M{H}_{pq}$, and $\M{P}$. This allows us to solve the identical \ac{QCQP} from \Cref{eq:QCQP_vswf_stiffness} now optimized exclusively over the physically realizable fields supported by a planar aperture.

\subsection{Comparison to Gaussian beam trap}

\begin{figure}[t!]
    \centering
    \includegraphics[width=\linewidth]{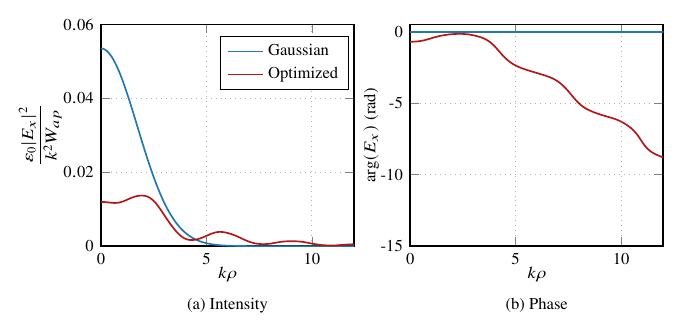}
    \caption{Radial profile of the transverse aperture field, with (a) normalized $|E_x|^2$, and (b) phase, $\arg\{E_x\}$.}
    \label{fig:aperture_fields}
\end{figure}

\begin{figure}[t]
    \centering
    \includegraphics[width=\textwidth]{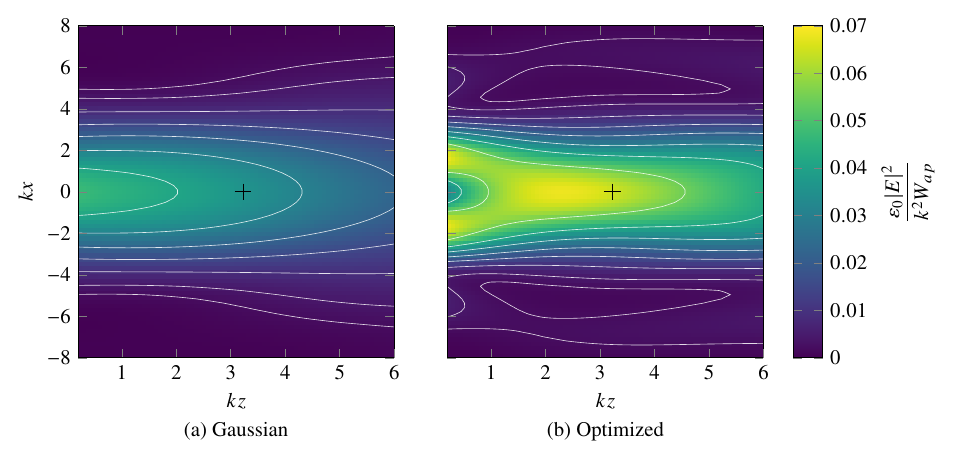}
    \caption{Normalized field intensity in the \(x\)-\(z\) plane for the (a) Gaussian and (b) (non-symmetric) optimized cases. The white curves denote iso-intensity contours and the black cross marks the focal point at \((kz,kx)=(3.23,0)\).}
    \label{fig:field-comparison}
\end{figure}

To demonstrate the practical utility of the proposed aperture formulation, the trapping performance of the optimized fields against that of a Gaussian reference beam is compared. We consider a lossless dielectric particle ($\phi=0$) with normalized volume~$\zeta=0.0719$, corresponding to the best trapping performance, and evaluated at the longitudinal position $kz \approx 3.23$. The focusing aperture has a dimensionless radius $kR=15$ and the Gaussian reference has the normalized beam waist $kw_0 \approx 3.45$. Both the Gaussian and optimized fields are expanded in the same discrete Bessel basis and are subject to the same aperture-energy constraint. To judge the trapping performance, dimensionless stiffness matrix $\hat{\V{H}}=\V{H}/(kW_\T{ap})$, where $W_\T{ap}$ is defined by \Cref{eq:aperture_intensity}, is used.

For a single-beam axisymmetric aperture, rotational symmetry about the optical axis forces the off-diagonal stiffness components to vanish at the equilibrium point, leaving a diagonal stiffness matrix whose entries are the principal confinement strengths.

An optical trap formed by a single Gaussian beam is inherently anisotropic. Radiation pressure introduces a preferred propagation direction, while linear polarization breaks rotational symmetry in the transverse plane. \Cref{tab:stiffness_comparison} compares this Gaussian reference with the corresponding optimized solutions.
For the considered parameter regime, the Gaussian trap is limited by its axial confinement, with longitudinal stiffness more than an order of magnitude smaller than the transverse components. Enforcing isotropy removes this bottleneck by balancing the principal curvatures, yielding a trap that is uniformly stable in all directions, at the cost of reducing the stronger transverse confinement. This isotropic optimum therefore provides the most robust measure of overall trap quality when stability against escape in any direction is required.

\begin{table}[htbp]
    \centering
    \small
    \caption{Comparison of the Gaussian reference and optimized solutions under aperture-energy normalization. More negative values correspond to stronger confinement. For the isotropic optimum, the components of the diagonal Hessian matrix correspond to the eigenvalues.}
    \begin{tabular}{lccc c}
        \hline
        Case & \(\hat{H}_{xx}\) & \(\hat{H}_{yy}\) & \(\hat{H}_{zz}\) & \(\T{Tr}(\hat{\V{H}})\) \\
        \hline
        Gaussian & \(-3.29\times10^{-2}\) & \(-5.35\times10^{-2}\) & \(-1.28\times10^{-3}\) & \(-8.76\times10^{-2}\) \\
        Isotropic optimum & \(-2.64\times10^{-2}\) & \(-2.64\times10^{-2}\) & \(-2.64\times10^{-2}\) & \(-7.93\times10^{-2}\) \\
        Anisotropic optimum & \(-1.28\times10^{-1}\) & \(-2.13\times10^{-1}\) & \(-2.46\times10^{-2}\) & \(-3.66\times10^{-1}\) \\
        \hline
    \end{tabular}
    \label{tab:stiffness_comparison}
\end{table}

Relaxing the isotropy constraint allows the optimization to prioritize stiffness in selected directions. The resulting anisotropic solution redistributes the available confinement, achieving significantly stronger stiffness along certain axes by exploiting the full directional freedom of the aperture, while leaving other directions weakly confined. As a result, it does not suppress the weakest axis and is therefore less robust against particle escape than the isotropic optimum. Such anisotropic traps are, nevertheless, of practical interest in scenarios where confinement is only required along specific directions, and controlled directional stiffness is more relevant than fully balanced confinement~\cite{Matheson2021}.
These results highlight the distinction between balanced and absolute trapping performance. For the aperture radius considered here, the optimal isotropic trap does not exceed the Gaussian reference in total trace, despite eliminating the axial weakness. In contrast, the anisotropic optimum significantly outperforms the Gaussian benchmark in trace. In general, minimizing the trace alone does not guarantee a stable trap, since one or more eigenvalues may remain positive. The corresponding optimized aperture profile is shown in \Cref{fig:aperture_fields}, while the resulting focal-plane field configurations are shown in \Cref{fig:field-comparison}. The white curves denote iso-intensity contours, and the black cross marks the equilibrium position at which the stiffness matrix is evaluated.

\begin{figure}[t!]
    \centering
    \includegraphics[width=0.50\linewidth]{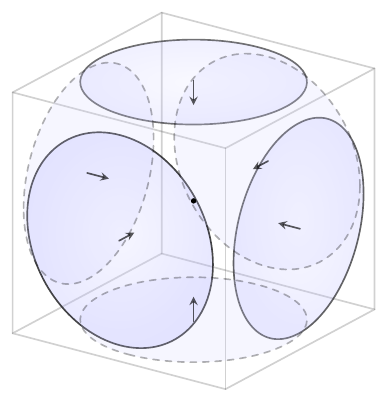}
    \caption{Schematic of the aperture configuration. Six circular apertures are placed on the faces of a cube, each with its local wave vector $k$ indicated. Every aperture supports two polarizations, represented by two independent sets of coefficients. The trapping point is shown as a black circle at the center.}
    \label{fig:Cube_aperture}
\end{figure}

\subsection{Comparison of Multi-Aperture and Vector Spherical Wave Approaches}
\label{sec:vsw_vs_ap}
To compare device-constrained illumination with the free-space optimum, we evaluate isotropic trapping bounds for a multi-aperture configuration and contrast them with the corresponding bounds obtained from a truncated \ac{VSWF} basis. The multi-aperture formulation naturally extends the single-aperture construction to several independently positioned and oriented lenses. Each aperture \(m\) is described by center position \(\V{t}_m\) and a local orthonormal frame whose longitudinal axis is aligned with the optical axis of the aperture. The total incident field is then obtained as the coherent superposition of the fields radiated by all apertures after a transformation from their local frames into the global trapping frame.
In the discrete setting, the aperture degrees of freedom are collected into a single stacked coefficient vector,
\begin{equation}
\M{a} = [\M a_1^\trans,\M a_2^\trans,\dots,\M{a}_m^\trans]^\trans .
\end{equation}
As the field at the trapping point depends linearly on these coefficients, all local quantities entering the force and stiffness, namely the electric field and its first and second spatial derivatives, also depend linearly on \(\M a\). Consequently, the force-balance constraints and the isotropic-stiffness objective retain the same quadratic structure as in \Cref{eq:QCQP_vswf_stiffness}, and the optimization remains a \ac{QCQP} in the global coefficient vector. The resulting matrices are generally dense, since they include all interference cross-terms between fields radiated by different apertures.

In the numerical implementation considered here, we employ two aperture configurations. First, a six-aperture arrangement with two orthogonal polarization channels per aperture is used, yielding twelve aperture subspaces in total. Second, a two-aperture arrangement with two orthogonal polarization channels per aperture is used, yielding eight aperture subspaces in total. In both configurations, the apertures are arranged in opposing pairs, and each local optical axis is directed toward the trapping center at the origin. The setup is schematically represented in~\Cref{fig:Cube_aperture}.

A meaningful comparison between the multi-aperture and \ac{VSWF} formulations requires a common quadratic normalization. Here, we use the normalization introduced in \Cref{apx:VSWF}, where the cycle-mean energy density is integrated over sphere $S_a$ of radius $a$, specifically, by
\begin{equation}
{W}_a = \dfrac{1}{4}\int\limits_{S_a} \left( \varepsilon |\V{E}|^2 + \mu |\V{H}|^2 \right)\T{d}S .
\end{equation}
This quantity provides a positive-definite quadratic measure of the electromagnetic field amplitude on the spherical boundary and serves as the common resource constraint for both formulations. The formulation in \acs{VSWF} acts as an aperture that circumscribes the trapping or tweezing point.

\begin{figure}[t!]
    \centering
    \includegraphics[width=\linewidth]{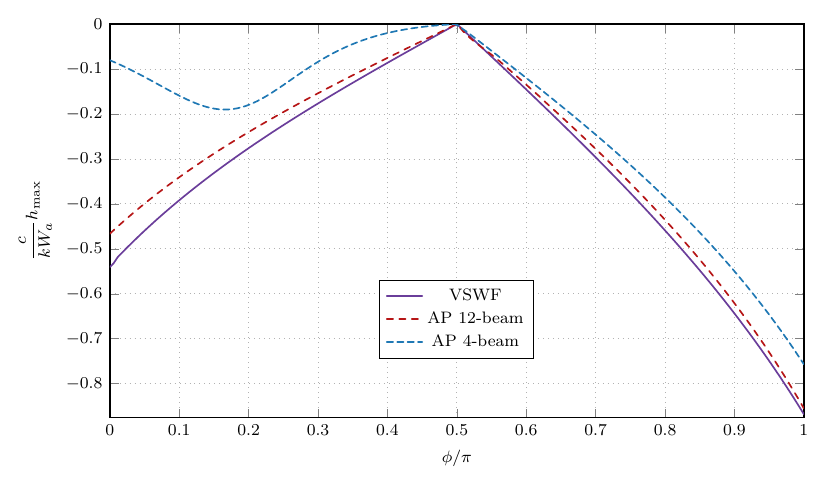}
    \caption{Optimal isotropic trapping stiffness $ch_{\max}/(k W_a)$ as a function of polarizability phase $\phi$ for the \ac{VSWF} formulation and the corresponding six-aperture and two-aperture subspaces, using common spherical normalization $W_a$.}
    \label{fig:vsw_vs_ap}
\end{figure}

\begin{figure}[t!]
    \centering
    \includegraphics[width=\linewidth]{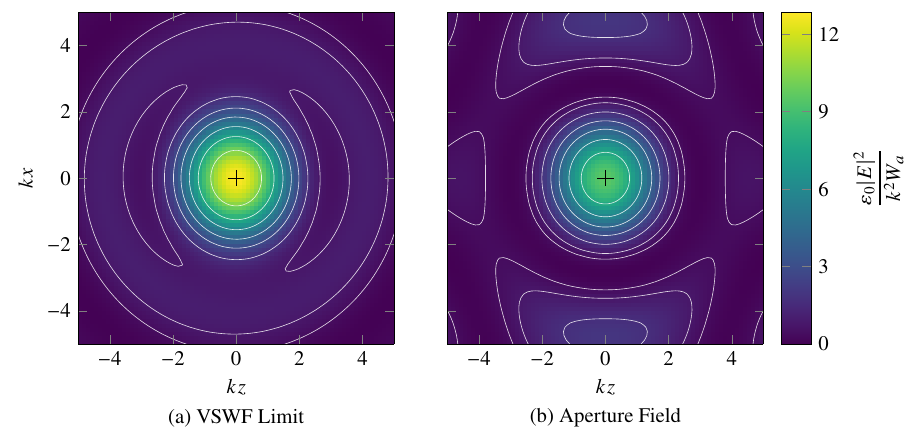}
    \caption{Normalized field intensity in the $x$-$z$ plane for representative optimal fields from the isotropic-stiffness optimization at $\phi=0$, obtained for the \ac{VSWF} formulation $(a)$ and the corresponding six-aperture subspace $(b)$. The white curves denote iso-intensity contours, and the black cross marks the focal point at the origin.}
    \label{fig:vsw_vs_ap_field}
\end{figure}

Using this common normalization, we compute the optimal isotropic trapping stiffness for both the \ac{VSWF} formulation and the corresponding multi-aperture subspaces as polarizability phase $\phi$ is varied. The resulting curves are shown in \Cref{fig:vsw_vs_ap}.  For all formulations, the optimal stiffness approaches zero at $\phi=\pi/2$, but for phases larger than $\pi/2$ the trap re-emerges and becomes stronger again.

The six-aperture subspace with two orthogonal polarization channels per aperture follows the \ac{VSWF} trend more closely because it spans a larger set of admissible field configurations, whereas the two-aperture subspace with two orthogonal polarization channels per aperture is more constrained. This reduced flexibility becomes especially visible near $\phi=0$, where the two-aperture solution departs from the other two curves. Of course, the precise shape of each curve depends on the specific parameter tuning, but the figure clearly shows that the dependence on phase is not trivial and that the trapping performance is strongly coupled to the available field degrees of freedom. In this regime, the optimizer can only use a limited combination of aperture modes, so the field cannot satisfy the force and isotropy constraints as effectively as in the larger subspaces.

Although the curves in \Cref{fig:vsw_vs_ap} appear different from those obtained under the previous normalization in \Cref{fig:pareto_plot}, this difference is largely an artifact of the chosen resource metric. In the present formulation, all fields are normalized by the common spherical quadratic measure \(W_a\), whereas the earlier results were normalized by incident power \(P_0\). If the current solutions were instead rescaled by their own \(P_0\), the phase dependence of optimal stiffness would recover the same general shape as in the previous formulation. The change of normalization, therefore, does not alter the underlying family of optimal fields, but it does change how their performance is weighted and compared within a fixed resource budget.

Finally, although the multi-aperture curve follows the same broad trend as the \ac{VSWF} curve, the remaining separation between them is physically meaningful. It reflects the fact that the finite-aperture system can access only the subset of spherical-wave excitations that can be synthesized by the chosen arrangement of lenses and polarizations. The \ac{VSWF} result should therefore be interpreted as the free-space upper bound under the common spherical normalization, while the multi-aperture result quantifies the extent to which that bound can be approached by the present device-constrained geometry.
A representative pair of optimal field distributions at $\phi=0$ is shown in~\Cref{fig:vsw_vs_ap_field} to illustrate the corresponding spatial structure of the free-space and multi-aperture optima. The VSWF solution exhibits a more concentrated and nearly radially symmetric focal profile, whereas the multi-aperture field retains the same central maximum but develops pronounced side lobes and a less uniform outer envelope, reflecting the reduced modal freedom of the finite-aperture geometry.

\section{Conclusion}
We have presented a general framework for determining fundamental bounds on optical force and trap stiffness under the dipole approximation of the response of the trapped particle. By expressing the local field and its spatial derivatives as linear functions of a chosen set of basis functions, both the optical force and the stiffness matrix can be written as Hermitian quadratic forms of the field coefficients. This allows us to formulate the optimal trapping and tweezing problems as a \acf{QCQP}, and to evaluate performance limits under prescribed physical constraints.

Using regular \acfp{VSWF}, we obtain the absolute free-space limits on the force and stiffness achievable by any monochromatic field. This formulation provides a natural description of the complete radiation space and allows the identification of the electromagnetic degrees of freedom that actively contribute to trapping. The analysis confirms that only a limited set of local field components and derivatives are relevant to dipole confinement, enabling us to reduce the optimization space without altering the underlying optimum.

To tighten the bounds and align them with practical settings, an aperture-based formulation was introduced in which the incident fields are restricted to those generated by finite circular apertures. This approach yields consistent bounds directly relevant to the implementation of a focused beam using lenses. Comparison with standard Gaussian field illumination shows that the optimized aperture fields can substantially reshape the stiffness and outperform the Gaussian reference.  At the same time, the isotropic formulation reveals the extent to which practical field synthesis is limited by finite aperture size and directional constraints.
The proposed framework can guide lens inverse design for better trapping and tweezing performance since it offers the optimal field at the aperture. 

We also presented a direct comparison between multi-aperture fields and the corresponding truncated \ac{VSWF} subspace under a common normalization. This comparison shows that the finite-aperture system reproduces the same broad qualitative dependence on the polarizability phase as the free-space optimum, while remaining separated from it because only a restricted subset of spherical-wave excitations can be realized by the chosen lens geometry.

Overall, this work establishes a rigorous framework for studying both free-space and device-constrained limits. Beyond the specific examples treated here, the same methodology can be extended to more general illumination geometries, alternative resource constraints, and more complex particle models, including magnetic, anisotropic, or multipolar responses, as long as the appropriate metrics can be cast into a quadratic form. In this sense, the present work provides both a rigorous benchmark for realistic optical tweezer design and a flexible theoretical basis for future investigations of the limits of field-matter interaction.

\appendix
\section{Vector Spherical Wave Functions \label{apx:VSWF}}

In this paper, the definition of vector spherical waves follows \cite{1988_Hansen_SVW, 2016_Kristensson_Scattering}. In this definition, the incident electromagnetic field described by fields~$\V{E}(\V{r}), \V{H}(\V{r})$ regular at the origin is expanded as 
\begin{equation}
\begin{aligned}
    \V{E}(\V{r}) &= k \sqrt{Z} \sum_{\nu} a_\nu \, \V{u}^{(1)}_\nu(k \V{r}) \\
    \V{H}(\V{r}) &= -\T{i} \dfrac{k}{\sqrt{Z}} \sum_{\nu} a_\nu \, \V{u}^{(1)}_{\bar{\nu}}(k \V{r}).
    \label{eq:VSWepand}
\end{aligned}
\end{equation}
Vector function~$\V{u}^{(1)}_\nu(k \V{r})$ is the regular spherical vector wave~\cite[Chap.~7]{2016_Kristensson_Scattering}, characterized by multi-index~$\nu$, which collects indices~$\tau$ (TE/TM wave),~$l \geq 1$ (integer spherical order),~$m = 0, \cdots, l$ (integer magnetic number), and~$\sigma$ (odd/even in spherical angle $\varphi$). The bar over the spherical index indicates a transition between TE and TM waves. Functions~$\V{u}^{(1)}_\nu(k \V{r})$ are divergence-free and orthogonal on any spherical surface centered at the origin. The validity of expansion~\Cref{eq:VSWepand} is given by the property
\begin{equation}
    \begin{aligned}
    \dfrac{1}{k} \nabla \times \V{u}^{(1)}_\nu(k \V{r}) = \V{u}^{(1)}_{\bar{\nu}}(k \V{r}) \\
    \dfrac{1}{k} \nabla \times \V{u}^{(1)}_{\bar{\nu}}(k \V{r}) = \V{u}^{(1)}_\nu(k \V{r})
    \end{aligned}
\end{equation}
built in the construction of functions~$\V{u}^{(1)}_\nu(k \V{r})$. The cycle-mean inward power passing through a surface enclosing the origin can be evaluated as~\cite{2016_Kristensson_Scattering}
\begin{equation}
    P_\T{in} = \dfrac{1}{8} \M{a}^\T{H} \M{a}.
\end{equation}
In the lossless background, the cycle-mean power
\begin{equation}
    P_\T{out} - P_\T{in} = \dfrac{1}{2} \T{Re} \oint\limits_S \left( \V{E} \times \V{H}^* \right) \cdot \T{d}\V{S}
\end{equation}
must vanish, and the same power~$P_\T{out} = P_\T{in}$ must flow outward from the enclosing surface in the absence of the scatterer.

If a scatterer is present, the total electromagnetic field outside its smallest circumscribing sphere can be written as
\begin{equation}
\begin{aligned}
    \V{E}(\V{r}) &= k \sqrt{Z} \sum_{\nu} a_\nu \, \V{u}^{(1)}_\nu(k \V{r}) + f_\nu \, \V{u}^{(4)}_\nu(k \V{r}) \\
    \V{H}(\V{r}) &= -i\dfrac{k}{\sqrt{Z}} \sum_{\nu} a_\nu \, \V{u}^{(1)}_{\bar{\nu}}(k \V{r}) + f_\nu \, \V{u}^{(4)}_{\bar{\nu}}(k \V{r}),
    \label{eq:VSWepand1}
\end{aligned}
\end{equation}
where functions~$\V{u}^{(4)}_\nu(k \V{r})$ are outgoing vector spherical waves and coefficients~$\M{f}$ represent the scattering typically described by~\cite{2016_Kristensson_Scattering} transition matrix~$\M{T}$ defined by relation~$\M{f} = \M{T} \M{a}$. In such a case, the cycle-mean power
\begin{equation}
    P_\T{out} - P_\T{in} = \dfrac{1}{2} \T{Re} \oint\limits_S \left( \V{E} \times \V{H}^* \right) \cdot \T{d}\V{S} = \dfrac{1}{2} \left( \M{f}^\T{H} \M{f} + \T{Re} \left[ \M{a}^\T{H} \M{f} \right] \right)
\end{equation}
gives minus the power absorbed in the scatterer and
\begin{equation}
    P_\T{s} = \dfrac{1}{2} \M{f}^\T{H} \M{f}
\end{equation}
gives the cycle-mean scattered power.

The cycle-mean energy density per unit length evaluated from the incident field in \Cref{eq:VSWepand} and integrated over the spherical surface $S_a$ of radius $a$, which is used in \Cref{sec:vsw_vs_ap}, can be written as
\begin{equation}
   \dfrac{1}{4} \oint\limits_{S_a} \left( \varepsilon \left| \V{E} \right|^2 + \mu \left| \V{H} \right|^2 \right) \T{d}S = \dfrac{ \left( kr \right)^2}{4 c} \sum\limits_\nu  \left| a_\nu \right|^2 \left[ \left| R_{1l}^{\left( 1 \right)} \left( kr \right) \right|^2 + \left| R_{2l}^{\left( 1 \right)} \left( kr \right) \right|^2 + \left| R_{3l}^{\left( 1 \right)}  \left( kr \right) \right|^2 \right]
\end{equation}
with $R_{\tau l}^{\left( 1 \right)}$ denoting the radial part~\cite[Chap.~7.2]{2016_Kristensson_Scattering} of function $\V{u}_\nu^{\left( 1 \right)}$ and where the orthonormality of vector spherical harmonics is employed over a spherical surface.

\section{Aperture Fields}
\label{appAperture}
Consider an aperture in the plane $z=0$, with a transverse electric field
\begin{equation}
    \V{E}_\perp(\V{r}_\perp,0)=E_\perp(\V{r}_\perp)\,\hat{\M{e}},
\end{equation}
linearly polarized along a fixed unit vector $\hat{\M{e}}\perp\hat{\V{z}}$. The transverse position vector in the aperture plane is denoted by~$\V{r}_\perp=x\hat{\V{x}}+y\hat{\V{y}}$.

The angular plane wave spectrum is defined as an integral over the aperture
\begin{equation}
    \tilde{\V{E}}_\perp(\V{k}_\perp, 0)=\frac{1}{2\pi}\int \limits_{\T{ap}} \V{E}_\perp(\V{r}_\perp,0)\mathrm{e}^{-\T{i}\V{k}_\perp\cdot\V{r}_\perp}\,\mathrm{d}^2\V{r}_\perp.
\end{equation}

In a source-free region, the total electric field must have vanishing divergence. The spectrum of the total electric field thus satisfies the transversality condition
\begin{equation}
    \V{k}\cdot\tilde{\V{E}} (\V{k}_\perp, 0)=0,
\end{equation}
where
\begin{equation}
    \V{k}=\V{k}_\perp+k_z\hat{\V{z}}, \qquad k_z=\sqrt{k^2-|\V{k}_\perp|^2}, \qquad \operatorname{Im}\{k_z\}\ge 0.
\end{equation}
Therefore, the longitudinal spectral component is
\begin{equation}
    \tilde{E}_z(\V{k}_\perp, 0)=-\frac{\V{k}_\perp\cdot\hat{\M{e}}}{k_z}\tilde{E}_\perp(\V{k}_\perp, 0).
\end{equation}
and the full electric spectrum can be written as
\begin{equation}
    \tilde{\V{E}}(\V{k}_\perp,0)=\tilde{E}_\perp(\V{k}_\perp)\left(\hat{\M{e}}-\frac{\V{k}_\perp\cdot\hat{\M{e}}}{k_z}\hat{\V{z}}\right).
\end{equation}
The propagated electric field in half-space~$z > 0$ then follows from the inverse angular-spectrum representation
% \begin{equation}
%     \V{E}(\V{r}_\perp,z)=\int \tilde{E}_\perp(\V{k}_\perp)\left(\hat{\V{e}}-\frac{\V{k}_\perp\cdot\hat{\V{e}}}{k_z}\hat{\V{z}}\right)\mathrm{e}^{\T{i}\V{k}_\perp\cdot\V{r}_\perp}\mathrm{e}^{\T{i}k_z z}\,\mathrm{d}^2\V{k}_\perp.
% \end{equation}
\begin{equation}
    \V{E}(\V{r}_\perp,z)= \frac{1}{2\pi}\int \tilde{\V{E}}(\V{k}_\perp,0)\T{e}^{\T{i}\V{k}_\perp\cdot\V{r}_\perp}\T{e}^{\T{i}k_z z}\,\T{d}^2\V{k}_\perp.
\end{equation}
To obtain the magnetic field, we use Faraday's law in the spectral domain,
\begin{equation}
    \tilde{\V{H}}(\V{k}_\perp,0)=\frac{1}{\omega\mu}\V{k}\times\tilde{\V{E}}(\V{k}_\perp,0).
\end{equation}
Substituting the electric spectrum yields
\begin{equation}
    \tilde{\V{H}}(\V{k}_\perp,0)=\frac{\tilde E_\perp(\V{k}_\perp)}{\omega\mu}\left[ \V{k}_\perp\times\hat{\M{e}} + \hat{\V{z}}\times\left(k_z\hat{\M{e}}+\frac{\V{k}_\perp\cdot\hat{\M{e}}}{k_z}\V{k}_\perp\right)\right].
\end{equation}
Accordingly, the propagated magnetic field for~$z > 0$ is
\begin{equation}
    \V{H}(\V{r}_\perp,z)=\frac{1}{2\pi} \int \tilde{\V{H}}(\V{k}_\perp,0)\mathrm{e}^{\T{i}\V{k}_\perp\cdot\V{r}_\perp}\mathrm{e}^{\T{i}k_z z}\,\mathrm{d}^2\V{k}_\perp.
\end{equation}
These expressions are general, valid for any linearly polarized transverse spectrum $\tilde{E}_\perp(\V{k}_\perp)$.

The time-averaged electromagnetic energy density per unit length in the plane, $z=0$, which is used in~\Cref{sec:aperture} and written there equivalently as an integral over $\T{d}S$, is
\begin{equation}
    W_\T{ap}=\frac{1}{4}\int\limits_\T{ap} \left(\varepsilon|\V{E}(\V{r}_\perp,0)|^2+\mu|\V{H}(\V{r}_\perp,0)|^2\right)\,\mathrm{d}^2\V{r}_\perp.
\end{equation}
By Parseval's theorem, this can be written in the spectral domain as
\begin{equation}
    W_\T{ap}=\frac{1}{4}\int\left(\varepsilon|\tilde{\V{E}}(\V{k}_\perp)|^2+\mu|\tilde{\V{H}}(\V{k}_\perp)|^2\right)\,\mathrm{d}^2\V{k}_\perp,
\end{equation}
where
\begin{equation}
\varepsilon|\tilde{\V{E}}(\V{k}_\perp)|^2=\varepsilon|\tilde E_\perp(\V{k}_\perp)|^2\left(1+\left|\frac{\V{k}_\perp\cdot\hat{\M{e}}}{k_z} \right|^2\right).
\end{equation}
and
\begin{equation}
    \mu|\tilde{\V{H}}(\V{k}_\perp)|^2=\frac{\varepsilon}{k^2}|\tilde E_\perp(\V{k}_\perp)|^2\left(|\V{k}_\perp|^2-|\V{k}_\perp\cdot\hat{\M{e}}|^2+\left|k_z\hat{\M{e}}+\frac{\V{k}_\perp\cdot\hat{\M{e}}}{k_z}\V{k}_\perp\right|^2\right).
\end{equation}
If the aperture field is axisymmetric, the transverse spectrum depends only on the radial component~$k_\rho=|\V{k}_\perp|$. Using polar coordinates in the transverse spectral plane, $\mathrm{d}^2\V{k}_\perp=k_\rho\,\mathrm{d}k_\rho\,\mathrm{d}\varphi_k$, the propagated electric field becomes
% \begin{equation}
%     \V{E}(\V{r}_\perp,z)=\int_0^\infty\int_0^{2\pi}\tilde{E}_\perp(k_\rho)\left(\hat{\V{e}}-\frac{\V{k}_\perp\cdot\hat{\V{e}}}{k_z}\hat{\V{z}}\right)\mathrm{e}^{\T{i}\V{k}_\perp\cdot\V{r}_\perp}\mathrm{e}^{\T{i}k_z z}\,k_\rho\,\mathrm{d}\varphi_k\,\mathrm{d}k_\rho.
% \end{equation}
\begin{equation}
\begin{aligned}
\V{E}_\perp(\V{r}_\perp,z) &= \hat{\M{e}} \int \limits_0^\infty  \tilde{E}_\perp(k_\rho) 
     \T{J}_0 \left( k_\rho \rho \right)
    \mathrm{e}^{\T{i}k_z z}
    \,k_\rho\,\mathrm{d}k_\rho \\
E_z(\V{r}_\perp,z) &= - \T{i} \left( \hat{\V{r}}_\bot \cdot \hat{\M{e}} \right) \int \limits_0^\infty  \tilde{E}_\perp(k_\rho) 
     \dfrac{k_\rho}{k_z}  \T{J}_1 \left(k_\rho \rho \right)
    \mathrm{e}^{\T{i}k_z z}
    \,k_\rho\,\mathrm{d}k_\rho \\
    \tilde{E}_\perp(k_\rho) &= \int E_\perp(\rho) \T{J}_0 \left( k_\rho \rho \right) \rho \T{d} \rho.
\end{aligned}
\end{equation}

By assuming a real-valued wavenumber, the expression for energy density per unit length reduces to the radial form
\begin{equation}
    W_\T{ap}=\dfrac{\pi\varepsilon}{2}\int \limits_0^\infty w(k_\rho)\,|\tilde{E}_\perp(k_\rho)|^2\,k_\rho\,\mathrm{d}k_\rho,
    \label{eq:WcycleMean}
\end{equation}
where
\begin{equation}
    w(k_\rho)=
    \begin{cases}
        1+\dfrac{k^2}{k_z^2}, & k_\rho\le k,\\[1ex]
        \dfrac{k_\rho^4}{k^2|k_z|^2}, & k_\rho>k.
    \end{cases}
\end{equation}

\begin{backmatter}
\bmsection{Funding}
This work was supported by the Czech Science Foundation under project No. 24-11678S. Martin Zlabek also acknowledges the support of the Czech Technical University in Prague under project No. SGS25/143/OHK3/3T/13. The work of Jakub Liska was supported by Natural Sciences and Engineering Research Council of Canada (NSERC) under Impact+ Research Training Award, the Québec ministry of economy, innovation and energy, and the Polytechnique C2MI research partnership, as well as additional benefits provided from their affiliations to the Regroupement Québécois sur les Matériaux de Pointes, \url{https://doi.org/10.69777/309032}, and the IVADO research consortium.

% \bmsection{Acknowledgment}

\bmsection{Disclosures}
The authors declare no conflicts of interest.

\bmsection{Data Availability Statement}
The data and code underlying the results presented in this paper are available in the GitHub repository \url{https://github.com/Zlabekma/Axi_aperture/tree/main}

% \bmsection{Supplemental document}

\end{backmatter}

\newpage
\bibliography{references,refOpticalManipulation,extraBib}

\end{document}